\documentclass[conference]{IEEEtran}
\IEEEoverridecommandlockouts
\usepackage{cite}
\usepackage{subfigure}
\usepackage{subcaption}
\usepackage{stfloats}
\usepackage{amsmath,amssymb,amsfonts}
\usepackage{hyperref}
\usepackage{amsthm}
\usepackage[ruled]{algorithm2e}	
\usepackage{graphicx}
\usepackage{textcomp}
\usepackage{xcolor}
\usepackage{booktabs}
\usepackage{float} 
\usepackage[justification=centering]{caption}
\def\BibTeX{{\rm B\kern-.05em{\sc i\kern-.025em b}\kern-.08em
    T\kern-.1667em\lower.7ex\hbox{E}\kern-.125emX}}
\usepackage{enumitem}

\newtheorem{assumption}{Assumption}

\begin{document}

\title{BLOCKS: Blockchain-supported Cross-Silo Knowledge Sharing for Efficient LLM Services }

\author{\IEEEauthorblockN{
Zhaojiacheng Zhou\IEEEauthorrefmark{2},
Hongze Liu\IEEEauthorrefmark{2},
Shijing Yuan\IEEEauthorrefmark{3},
Hanning Zhang\IEEEauthorrefmark{2},
Jiong Lou\IEEEauthorrefmark{2},
Chentao Wu\IEEEauthorrefmark{2},
Jie Li\IEEEauthorrefmark{4}\IEEEauthorrefmark{1}
}
\IEEEauthorblockA{
$\IEEEauthorrefmark{2}$Department of Computer Science and Engineering,
Shanghai Jiao Tong University,
China\\
}\
\IEEEauthorblockA{
$\IEEEauthorrefmark{3}$China Telecom Research Institute, Shanghai, China\\
}
\IEEEauthorblockA{ 
$\IEEEauthorrefmark{4}$
MoE Key Lab of Artificial Intelligence, Shanghai Jiao Tong University
}
\IEEEauthorblockA{ 
Email: 
\IEEEauthorrefmark{1}\{zzjc123, seniordriver233, 2019ysj, zhang5026871, lj1994, wuct, lijiecs\}@sjtu.edu.cn,
}
\thanks{\IEEEauthorrefmark{1} Jie Li is the corresponding author.}
\thanks{The code is in: https://anonymous.4open.science/r/BLOCKS.}
}

\maketitle

\begin{abstract}
The hallucination problem of Large Language Models (LLMs) has increasingly drawn attention. Augmenting LLMs with external knowledge is a promising solution to address this issue. However, due to privacy and security concerns, a vast amount of downstream task-related knowledge remains dispersed and isolated across various "silos," making it difficult to access. To bridge this knowledge gap, we propose a blockchain-based external knowledge framework that coordinates multiple knowledge silos to provide reliable foundational knowledge for large model retrieval while ensuring data security. Technically, we distill knowledge from local data into prompts and execute transactions and records on the blockchain. Additionally, we introduce a reputation mechanism and cross-validation to ensure knowledge quality and provide incentives for participation. Furthermore, we design a query generation framework that provides a direct API interface for large model retrieval. To evaluate the performance of our proposed framework, we conducted extensive experiments on various knowledge sources. The results demonstrate that the proposed framework achieves efficient LLM service knowledge sharing in blockchain environments.

\end{abstract}


\section{Introduction}
Large Language Models (LLMs) demonstrate impressive human-like text generation capabilities but suffer from notable limitations such as hallucination—the generation of plausible yet factually incorrect content~\cite{patil2024review}. A promising direction to mitigate this issue involves augmenting LLMs with external knowledge, which can enhance their accuracy and reliability. Techniques such as Retrieval-Augmented Generation (RAG)~\cite{zhao2024retrieval} and Chain of Knowledge (CoK)~\cite{li2024chain} have been proposed to incorporate such knowledge effectively.

Most existing research on LLM augmentation focuses on improving retrieval strategies, typically assuming access to centralized, static datasets~\cite{li2024chain}. However, emerging studies suggest that many LLM applications require multi-domain knowledge to achieve optimal performance~\cite{li2024chain, song2025injecting}. In practice, task-relevant knowledge is often dispersed across numerous isolated silos, making it challenging to access due to privacy concerns, competitive interests, and insufficient incentives for sharing. Although some systems can extract valuable information from the web~\cite{lai2024autowebglm}, the absence of a systematic cross-silo knowledge-sharing mechanism significantly limits their scalability and effectiveness. This issue is especially pronounced for comprehensive tasks requiring high-value, domain-specific knowledge, which remains largely underutilized.

Unifying siloed knowledge to produce coherent and actionable inputs for LLMs presents three primary challenges: \textbf{P1: Incentivization}, \textbf{P2: Quality of Service (QoS)}, and \textbf{P3: Security}. For \textbf{P1}, motivating silo owners to contribute knowledge requires addressing participation overhead and privacy concerns. For \textbf{P2}, integrating heterogeneous and unstructured data into LLM-compatible formats must preserve both data utility and privacy. Direct sharing of raw data poses privacy risks and may degrade LLM performance due to redundancies and structural inconsistencies, which also contribute to increased inference latency~\cite{zhao2024retrieval}. Moreover, LLM services are often delay-sensitive, while knowledge sharing is susceptible to latency stemming from network dynamics. For \textbf{P3}, the inherently untrusted nature of distributed contributors exposes the system to threats such as prompt injection attacks~\cite{yi2023benchmarking}.


Blockchain technology offers a promising foundation for addressing these challenges due to its inherent transparency, immutability, and traceability~\cite{TMC24maa2c,yuan2022jora,yuan2022jira}. By leveraging consensus protocols and on-chain reputation mechanisms, blockchain can provide decentralized incentives and secure interactions among participants~\cite{nguyen2025blockchain}. However, prior efforts—such as Collaborative Databases~\cite{li2024coraldb} and Distributed Knowledge Sharing Systems~\cite{li2020preserving}—do not target LLM-specific knowledge integration and often incur performance bottlenecks due to high data transmission overheads. Moreover, most of the reputation-based systems adopt open-loop architectures, which leave critical components outside the scope of monitoring, thereby introducing security vulnerabilities~\cite{khezr2022edge,yang2024secure}.

To this end, we propose a \textbf{BLO}ckchain-based \textbf{C}ross-silo \textbf{K}nowledge-\textbf{S}haring framework (\textbf{BLOCKS}), built on the COSMOS~\cite{kwon2019cosmos}, designed to coordinate multiple silos and deliver trustworthy, domain-specific knowledge for LLM services. To address \textbf{P1} (incentive) and \textbf{P3} (security), we implement a smart contract-based reputation mechanism that ensures data quality while incentivizing participation. To further safeguard the integrity of shared knowledge, we introduce a benchmark-based validation process that penalizes contributors whose responses fall below a predefined similarity threshold. For \textbf{P2} (quality of service), BLOCKS facilitates secure knowledge distillation by converting local data into prompts, which are then transacted and recorded on-chain. Additionally, we integrate a caching mechanism, \textbf{PROCache}, to enhance the efficiency of frequently accessed prompt retrieval. To further optimize blockchain ledger storage and parallel request handling, we implement a hash bucket key-value store layered over the IAVL-Tree, storing knowledge filtered and prioritized by \textbf{PROCache}. To the best of our knowledge, this is the first work to construct a blockchain-based external knowledge framework tailored for LLMs. Our key contributions are summarized as follows:
\begin{itemize}[leftmargin=*]
    \item We propose \textbf{BLOCKS}, a blockchain platform for distributed knowledge sharing, which integrates \textbf{PROCache} to improve prompt retrieval efficiency.
    \item We introduce a smart contract-based reputation mechanism to enforce security and incentivize honest participation.
    \item We evaluate BLOCKS across diverse knowledge sources and demonstrate its effectiveness in enabling secure and efficient LLM augmentation.
\end{itemize}

\noindent\textbf{Paper Organization:} Section~\ref{sec:motivation} outlines the motivation behind our work. Section~\ref{Implement} introduces the design of \textsc{Blocks}. Section~\ref{Overview} details the implementation. Section~\ref{Experiment} presents the experimental results.Section~\ref{Conclusion} concludes the paper.

\section{Motivation}
\label{sec:motivation}

\subsection{Multi-Source Knowledge for LLM Services}

Integrating external knowledge is pivotal for enhancing the performance of Large Language Models (LLMs), particularly when tackling complex downstream tasks and mitigating issues such as hallucination. A growing body of research has shown that incorporating diverse sources of knowledge significantly strengthens the capabilities of LLMs~\cite{wang2025bring,zhao2024retrieval,liu2024llm,li2024chain}. In particular, recent findings suggest that exposing LLMs to multi-source and multidisciplinary knowledge leads to marked improvements in performance, especially on challenging tasks~\cite{li2024chain}.

\begin{figure}[t!]
    \centering
    \includegraphics[width=3in]{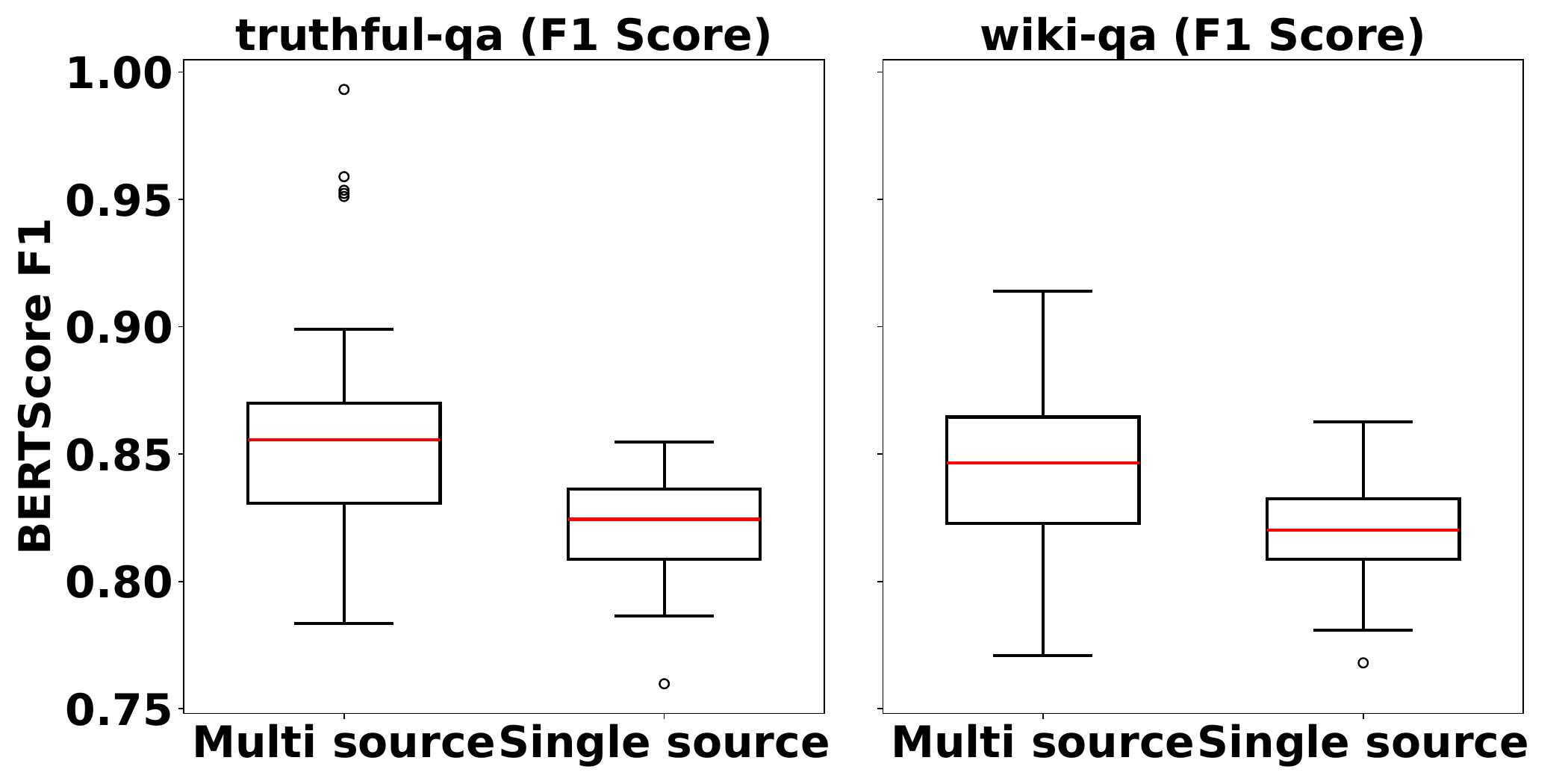}
    \caption{\small LLM performance improvement via multi-source knowledge integration.}
    \label{fig:multi-source}
\end{figure}

To empirically demonstrate this, we conduct a preliminary experiment using the widely adopted BERTScore metric~\cite{zhang2019bertscore} to compare LLM performance when leveraging single-source versus multi-source knowledge. The evaluation is performed on questions from the \textit{TruthfulQA}~\cite{lin2022truthfulqameasuringmodelsmimic} and \textit{WikiQA}~\cite{yang-etal-2015-wikiqa} datasets. Results, presented in Fig.~\ref{fig:multi-source}, show that multi-source knowledge provision consistently outperforms single-source configurations.

Despite its benefits, aggregating multi-source knowledge introduces significant challenges, as relevant information is typically dispersed across different entities~\cite{li2024coraldb,li2020preserving}. This highlights the pressing need for platforms that enable efficient, secure, and privacy-preserving access to decentralized knowledge.

\subsection{Blockchain-Based Knowledge Sharing Framework}

In our proposed framework, four key roles collaborate within the blockchain ecosystem~\cite{khezr2022edge,yang2024secure}:\\
\noindent
\textbf{Knowledge Providers}: Nodes responsible for generating prompts in response to LLM service requests.\\
\noindent
\textbf{LLM Servers}: Nodes that request external knowledge to satisfy user queries.\\
\noindent
\textbf{Trusted External Storage}: Components that archive historical knowledge and facilitate efficient retrieval.\\
\noindent
\textbf{Validators}: Nodes that verify the authenticity and relevance of provided prompts, while evaluating provider reputation.

Blockchain's decentralized and immutable nature offers a robust foundation for trustworthy and efficient knowledge sharing~\cite{li2024coraldb,li2020preserving}. While prior work has explored blockchain-based systems for knowledge exchange across domains, integrating such mechanisms into LLM services introduces several domain-specific challenges, necessitating a comprehensive design tailored for LLM adaptation.

\textbf{Incentivization}: Providing downstream knowledge imposes computational and communication overhead, making it essential to implement incentive mechanisms that fairly reward contributors. Smart contracts can automate compensation based on the quality of contributions, encouraging consistent and meaningful participation~\cite{nguyen2025blockchain}.

\textbf{Quality of Service (QoS)}: Ensuring timely responses to LLM queries is vital, particularly under fluctuating network conditions that may hinder data owner responsiveness. Moreover, integrating heterogeneous and unstructured knowledge into LLM-compatible formats must preserve data utility while protecting user privacy. Direct sharing of raw data not only risks privacy breaches but may also degrade model performance due to redundancies and inconsistencies, ultimately increasing inference latency~\cite{zhao2024retrieval}.

\textbf{Security}: Ensuring the integrity and authenticity of shared knowledge is critical, especially against threats like malicious prompt injection, such as context confusion attacks. The inherently untrusted nature of decentralized contributors in existing frameworks heightens vulnerability to such attacks~\cite{yi2023benchmarking}. Although previous works have introduced reputation systems to address these issues~\cite{khezr2022edge,yang2024secure}, many adopt open-loop designs that exclude essential components from continuous monitoring, thereby leaving security gaps unaddressed.

\section{BLOCKS System Design}
\label{Implement}

In this section, we introduce the interaction between a LLM service with a single raw data owners including how LLM generate query for the request of external knowledge and how data owner generate necessary knowledge without privacy leaking and data exposure. 

\subsection{Reputation Mechanism against Prompt Attack}
\label{sc:reputation}

As LLM systems become equipped with more plugins or access patterns, the downstream risks of prompt injection attack become elevated\cite{yi2023benchmarking}. To leverage the distributed network, we can implement cross validation for the detection of prompt injection attack. To further improve the robustness of system, we introduce a threshold detection for validators to accelerate the reputation dropping of malicious nodes. When validators start to validate the knowledge, the chain will ask a trusted service use low-cost model to evaluate the quality of knowledge and generate a threshold. If the result is out of the threshold, then we directly recognize this node as a malicious node and apply penalty.

Since a blockchain platform involves multiple participants with distinct roles, we design a reputation system tailored for each role. All reputation values are stored in trusted external storage and can only be modified through smart contracts.

To ensure reliable reputation evaluation, we leverage the metric of consistency (\(CS\)), based on the assumption that the expected value claims from honest participants should be consistent. However, due to individual differences in capability, slight variations in reported values may occur. The consistency metric measures whether a validator’s assessment aligns with other validators for the same prompt:

\vspace{-5pt}
\begin{equation}
    CS = \frac{\sum ||V - V_i|| \cdot R_{vi}}{(n-1) \cdot ||V_{\max} - V_{\min}|| \cdot R_{\max}}
    \label{eq:consistency}
\end{equation}
\vspace{-5pt}

where $V$ denotes the score assigned by a validator, and $V_i$ represents the score given by validator $i$. $R_{vi}$ indicates the reputation of validator $i$. $V_{\max}$, $V_{\min}$, and $R_{\max}$ refer to the maximum and minimum scores assigned by validators and the maximum validator reputation, respectively. We use $m$ and $n$ to represent the number of validations and the number of LLM services, respectively. 

\begin{figure}[t!]
    \centering
    \includegraphics[width=3in]{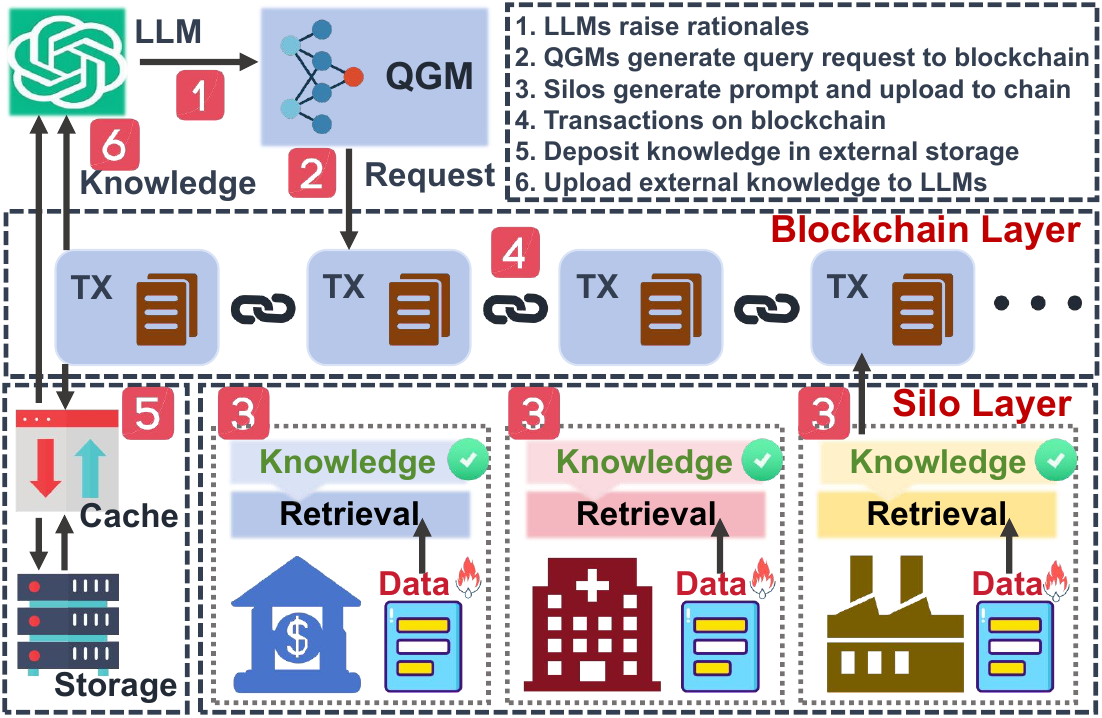}
    \caption{\small The System Overview}
    \label{fig:working flow}
\end{figure}

Considering the impact of the number of validations, we introduce the confidence metric (\(CF\)) to quantify the risk as the standard deviation (\(\text{STD}\)) of \(V\):

To ensure the credibility of knowledge, validators detect prompt attacks to safeguard knowledge quality from malicious agents and low-quality providers. The validation output \(V\) is quantified within the range \([0,1]\). The reputation for each role can be calculated as follows:
\begin{align}
    R_l &= \alpha \cdot (1 - \text{Mean}(CS))+(1-\alpha)\cdot R_l, \label{eq:LLM_reputation} \\
    R_p &= \frac{R_k + n \cdot \text{Mean}( V_i \cdot R_i) + m \cdot \text{Mean}(\text{Acc} \cdot R_l)}{n+m+1}- CF, \label{eq:prompt_reputation} \\
    R_k &= \alpha \cdot \text{Mean}(R_p)+(1-\alpha) \cdot R_k, \label{eq:provider_reputation} \\
    R_v &= \alpha \cdot (1 - \text{Mean}(CS))+(1-\alpha) \cdot R_v, \label{eq:validator_reputation1}
\end{align}

The reputations of LLM services ($R_l$), prompts ($R_p$), knowledge providers ($R_k$), and validators ($R_v$) are interdependent and updated dynamically. Using an exponential moving average with coefficient $\alpha$, reputations are computed based on performance consistency, feedback, and cross-validation, as defined in eqs~\(\ref{eq:LLM_reputation}\)--\(\ref{eq:validator_reputation1}\).

To ensure the credibility of knowledge, the permanence of reputation, and to provide sufficient incentives for distributed knowledge validation, we introduce the \textit{Proof of Impact} (PoI) consensus mechanism. In this protocol, the next block is proposed by the knowledge provider that has made the highest investment in knowledge quality. Rewards $C$ are distributed among all contributing knowledge providers in proportion to their impact, which is defined as the product of their reputation $R$ and the number of accesses $A$ to their associated prompts:

\begin{equation}
    C = R \cdot (\beta A_p + (1 - \beta) A_v),
\end{equation}

where $\beta \in [0,1]$ is the coefficient that balances the contributions between prompt generation $A_p$ and validation $A_v$.

Unlike Proof of Work (PoW), which is often criticized for resource inefficiency, PoI utilizes computational resources for meaningful validation tasks. It ensures an efficient, purpose-driven consensus process by monitoring each node’s influence—measured by the extent to which its contributed prompts are accessed by other nodes.

\subsection{Knowledge Cache Design}

Blockchain platforms are inherently limited in throughput, making it difficult to provide real-time external services for LLMs. To enhance efficiency and enable timely responses, external storage systems are commonly used to complement the blockchain \cite{10.14778/3476249.3476283}. Traditional cache management strategies such as LRU-k and LFU \cite{panda2016survey} focus on access frequency but may fail in adversarial environments. For example, malicious nodes could repeatedly access low-quality prompts, causing them to persist in the cache and potentially amplifying their negative impact.

To address this, we propose a \textbf{PR}iority \textbf{O}riented \textbf{Cache (PROCache)} strategy that integrates prompt reputation, value, and cost into cache prioritization.  PROCache consists of two main components: the Retrieve function and the cache storage system. The Retrieve function, implemented using LlamaIndex\cite{Liu_LlamaIndex_2022}, supports prompt embedding, indexing, and retrieval based on both metadata and content. The cache storage includes a history queue, a priority queue, and a key-value store. The smallest unit of storage in PROCache is a \textit{cache node}, defined by a node ID, content, metadata, and priority.

When a cache miss occurs, only the prompt's hash, access count, and a lightweight history record are stored in the history queue to minimize space usage. A prompt is added to the cache only after being accessed $k$ times in the history queue. Upon reaching this threshold, a full cache node is created, including the prompt content, metadata, reputation, and priority. The node ID is derived from the prompt's hash and count, and it is stored as a key-value pair in the main cache.

Eviction in PROCache is managed using a priority queue, where prompts are ordered based on their computed priority scores. This design ensures that PROCache retains high-quality, high-value prompts while mitigating cache pollution caused by malicious activity. To comprehensively account for access frequency, prompt cost, and storage size, the priority score of a prompt is defined as:

\begin{equation}
    \text{Priority} = \left( \frac{\text{Frequency} \times \text{Cost}}{\text{Size}} \right)^{R_t - R_b}
\end{equation}

Here, $R_b$ is a reputation threshold used to distinguish potentially malicious responses, and $R_t$ is the current reputation score of the prompt. This formulation ensures that prompts are promoted within the cache not merely by repeated access, but through a balanced assessment of quality and utility, making the system more resilient to adversarial manipulation.

\subsection{Transaction Design}

A complete BLOCKS transaction cycle comprises four sub-transactions involving the user, cache, supplier, and validator.
All temporary data of sub-transactions are encapsulated in a unified structure called a \textit{query-session}, uniquely identified by a session index. For persistent storage, the blockchain records key-value pairs using the COSMOS IAVL tree, leveraging different storage prefixes to distinguish between data types. The implementation defines four key-value pair types:\\
\noindent
\textbf{DataTable:} Stores the user prompt with its SHA-256 hash and an additional \texttt{hash count} to differentiate collisions. Each entry includes the prompt content and its supplier.\\
\noindent
\textbf{ReputationPrompt:} Uses the same hash and count as \texttt{DataTable}. Each record contains the prompt's reputation and history validations, where each validation includes the validator ID, the validator's reputation at the time, and the score assigned to the prompt.\\
\noindent
\textbf{ReputationSupplier:} Indexed by supplier ID, this entry holds the supplier’s current reputation score.\\
\noindent
\textbf{ReputationValidator:} Indexed by validator ID, this stores the validator’s reputation score.

This storage structure efficiently reduces large data redundancy through hashed indexing, ensures fast retrieval with short keys, and successfully handles hash collisions via an additional count. It ensures compactness, performance, and consistency within the COSMOS IAVL-Tree storage.

\section{Put Everything Together}
\label{Overview}

\subsection{BLOCKS Workflow}

The BLOCKS framework follows a four-stage transaction processes: \textit{Create Session}, \textit{Post Cache}, \textit{Update Knowledge}, and \textit{Update Validation}.

\noindent\textbf{1. Create Session}: A user submits a \texttt{create-session} transaction with their ID, query, and payment. The LLM generates a rationale using the Query Generation Module (QGM)~\cite{li2024chain} and record it in the session. Upon user identity verification, the blockchain broadcasts a \texttt{new-query} event and deducts the payment using the COSMOS bank module.

\noindent\textbf{2. Post Cache}: Cache services listening for \texttt{new-query} events respond via \texttt{post-cache} transactions with cached content and a hit flag. On a hit, the blockchain stores the content and broadcasts a \texttt{validate-update} event. On a miss, it broadcasts a \texttt{knowledge-update} event requesting input from knowledge suppliers.

\noindent\textbf{3. Update Knowledge}: Suppliers respond to \texttt{knowledge} \texttt{-update} events with \texttt{update-knowledge} transactions containing session index, supplier ID, and retrieved content. After enough submissions, the blockchain initiates validation via \texttt{update-validation} events.

\noindent\textbf{4. Update Validation}: Validators submit assessments through \texttt{update-validation} transactions. Feedback from official and regular validators is stored separately. Once validation is complete, the blockchain finalizes the session, broadcasts the result, updates reputations, and moves session data to persistent storage.

\begin{figure}[t]
    \centering
    \includegraphics[width=3in]{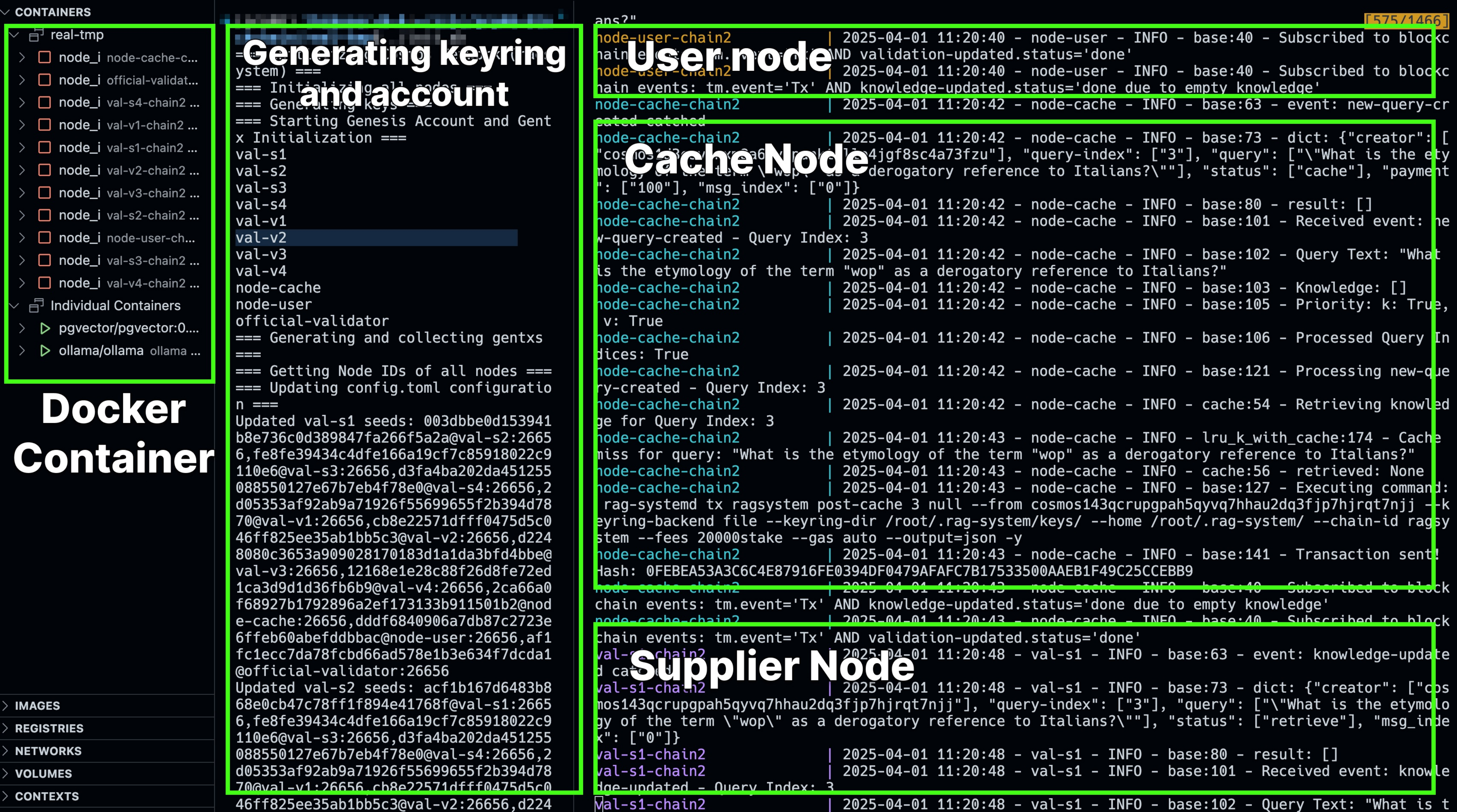} 
    \caption{The Implementation of BLOCKS}
    \label{fig:BLOCKSimplement}
\end{figure}

Figure~\ref{fig:working flow} illustrates the interaction of LLMs, cache services, knowledge providers, and validators. Validated knowledge is cached via PROCache, improving response efficiency. Fig.~\ref{fig:BLOCKSimplement} shows our device-level implementation.

For cached resales, external storage receives a service fee before distribution among providers. The reward $U_p$ distribution follows providers' reputation scores $R_p$, given by:
\begin{equation}
    U_p = R_{b} \times \frac{R_p}{\sum R_p}
\end{equation}

\subsection{Security Analysis}
\label{sc:secure}

We consider a Byzantine adversary capable of corrupting a fraction of nodes in the system, where the number of compromised nodes is bounded by $f < \frac{n}{3}$. Under this assumption, each request is expected to receive at most one-third of responses from malicious entities.

This work examines several adversarial strategies, some of which are inspired by prior research \cite{10.1145/3286978.3286981}. The primary attack vectors are outlined as follows:\\
\noindent
\textbf{Self-Promotion:} A malicious node engages in dishonest behavior to artificially inflate its own reputation score, thereby gaining undue influence in the system.\\
\noindent
\textbf{Collusion:} A group of malicious nodes collaborates to collectively boost their reputation scores, thereby increasing their dominance over the system.\\
\noindent
\textbf{Slandering:} A compromised leader node attempts to degrade the reputation scores of honest validators, undermining the trust and integrity of the system.

\begin{assumption}
Malicious prompts can be identified and receive lower reputation scores compared with honest prompts when evaluated by honest validators.
\end{assumption}

\begin{assumption}
Malicious validators do not possess an excessive advantage in validation compared to honest validators. Formally, the expected deviation in reputation assessment satisfies:
\begin{equation}
    E(|R_p^*, R_p^v|) \leq E(|R_p^*, R_p^m|)
\end{equation}

where $R_p^*$ is the true reputation of prompt $p$, $R_p^v$ is the reputation assigned by an honest validator, and $R_p^m$ is the reputation assigned by a malicious validator.
\end{assumption}

\subsubsection{Security of Reputation Mechanism}

Since the PoI mechanism directly relies on reputation, we first analyze the security of the reputation mechanism. 

As shown in Eq.~\eqref{eq:LLM_reputation} and Eq.~\eqref{eq:validator_reputation1}, both malicious users who provide distorted feedback and validators who perform incorrect validation suffer a decrease in reputation over time, ultimately reaching zero as iteration $t \to \infty$ :
\begin{equation}
    \lim_{t \to \infty} R_l^t = 0, \quad \lim_{t \to \infty} R_v^t = 0.
\end{equation}

\subsubsection{Security of Proof of Impact}
We analyze the security of PoI in two parts.\\
\noindent
\textbf{1. Contribution Calculation:}  
The total reward for a knowledge provider $k$ is:
\begin{equation}
    W_k^t = C_k^t + \sum_{p \in P_k} A_p^t R_p^t.
\end{equation}
Since $R_p^t$ converges to $V_p^*$ and malicious providers experience reputation degradation, it follows that:
\begin{equation}
    \lim_{t \to \infty} W_{k_m}^t = 0, \quad \text{for all malicious providers.}
\end{equation}
\textbf{2. Stability of Reputation Equilibrium:}  
Over infinite validation rounds, assuming honest validators dominate, the system stabilizes as:
\begin{equation}
    \forall p, \quad \lim_{t \to \infty} R_p^t = V_p^*.
\end{equation}
Thus, only high-reputation providers receive rewards, deterring malicious behavior.

\section{Experiment}
\label{Experiment}

\begin{figure*}[htp]
    \centering
    \begin{minipage}[t]{0.24\linewidth}
        \centering
        \includegraphics[width=\linewidth]{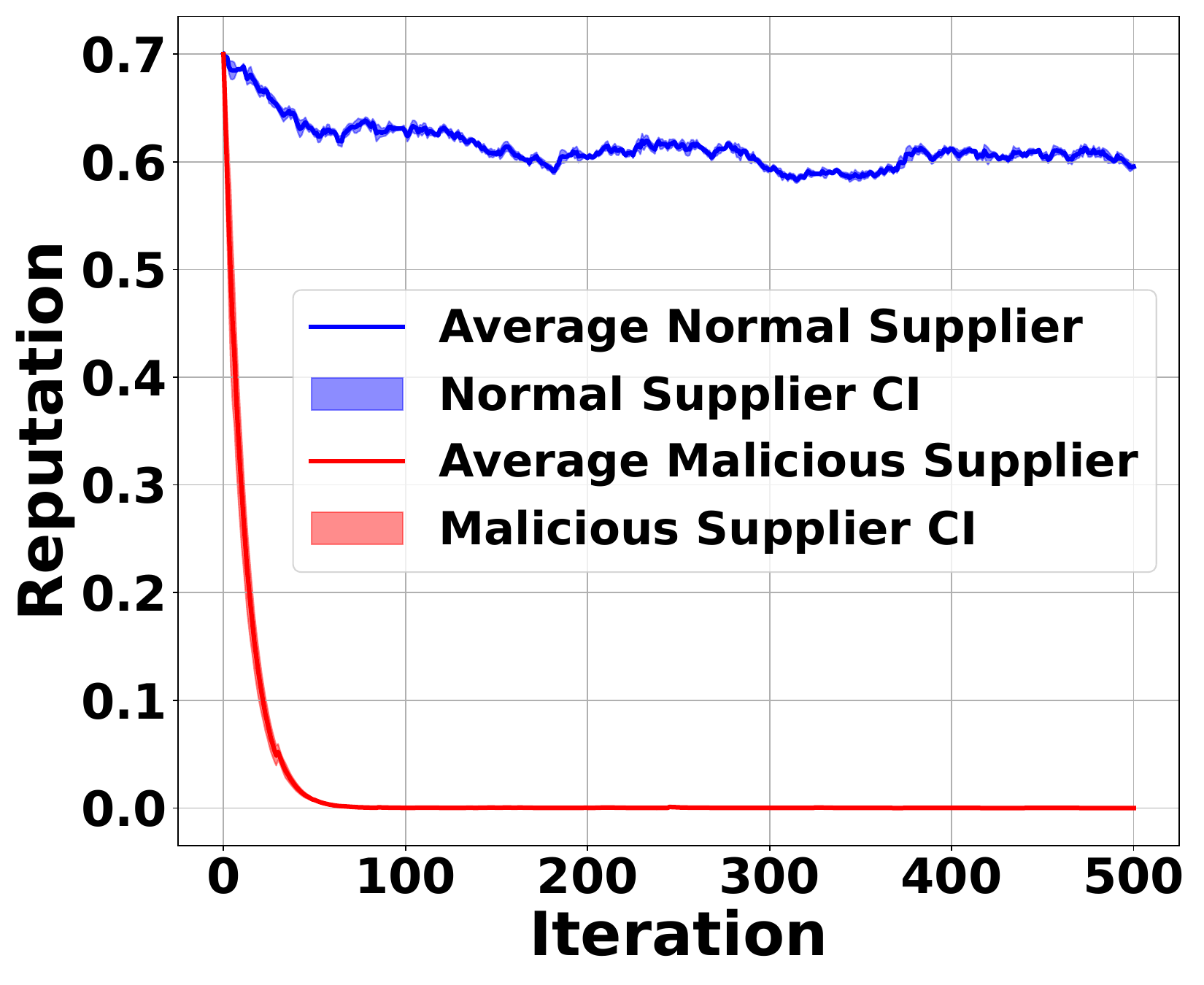}
        \caption{Supplier Reputation}
        \label{fig:RP}
    \end{minipage}
    \begin{minipage}[t]{0.24\linewidth}
        \centering
        \includegraphics[width=\linewidth]{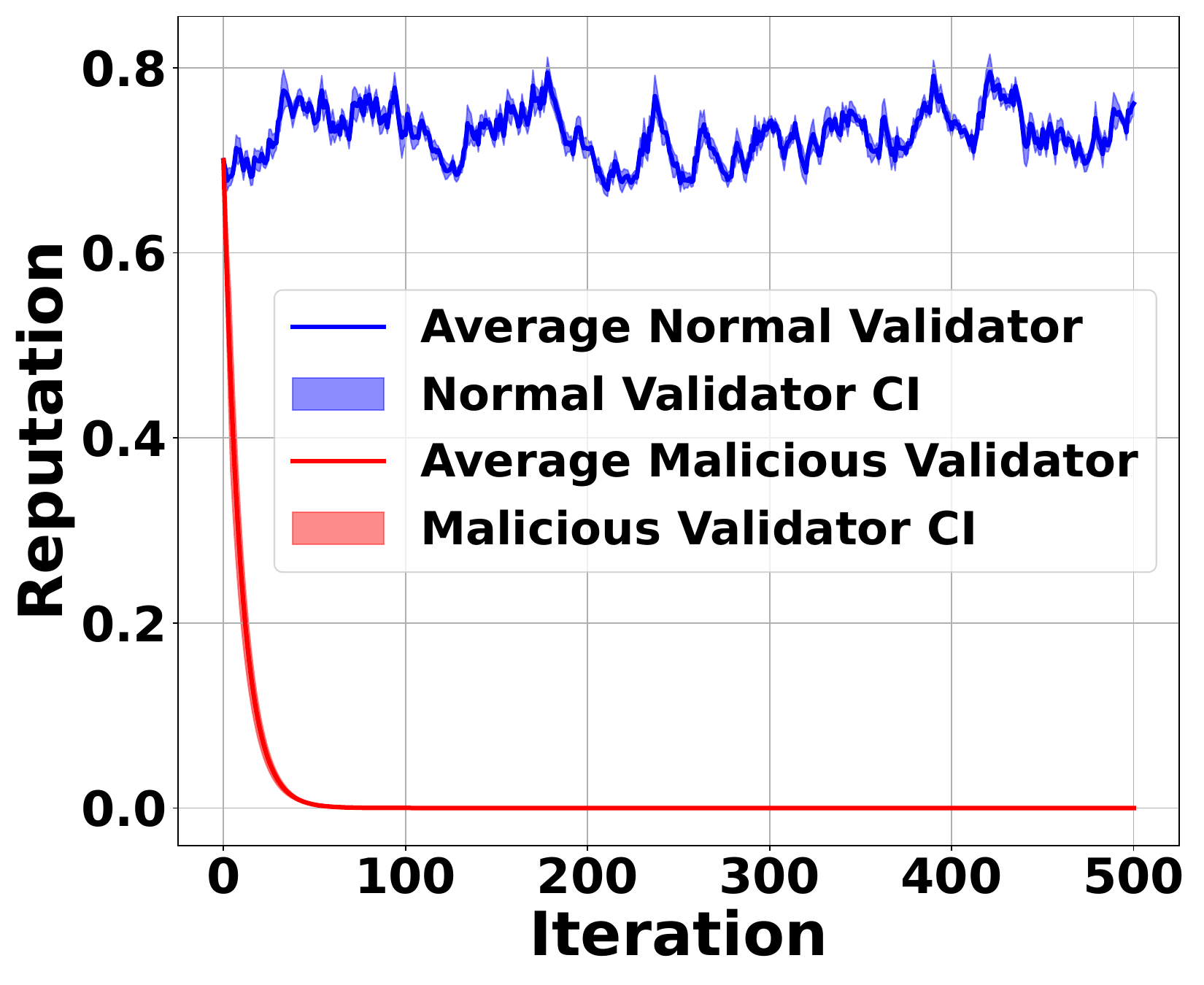}
        \caption{\small Validator Reputation}
        \label{fig:RV}
    \end{minipage}
    \begin{minipage}[t]{0.27\linewidth}
        \centering
        \includegraphics[width=\linewidth]{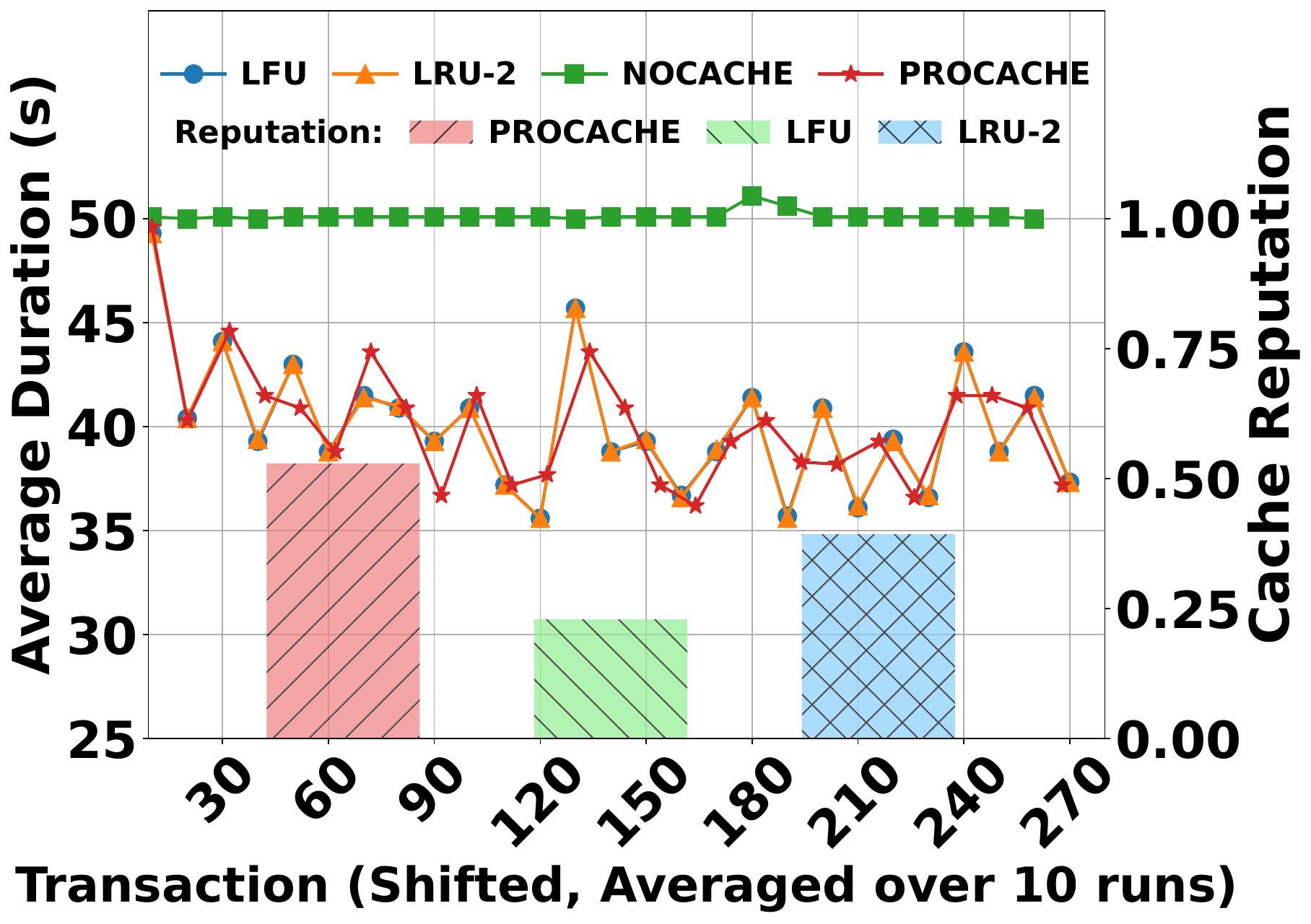}
        \caption{\small PROcache QoS Performance}
        \label{fig:Cache}
    \end{minipage}
    \begin{minipage}[t]{0.23\linewidth}
        \centering
        \includegraphics[width=\linewidth]{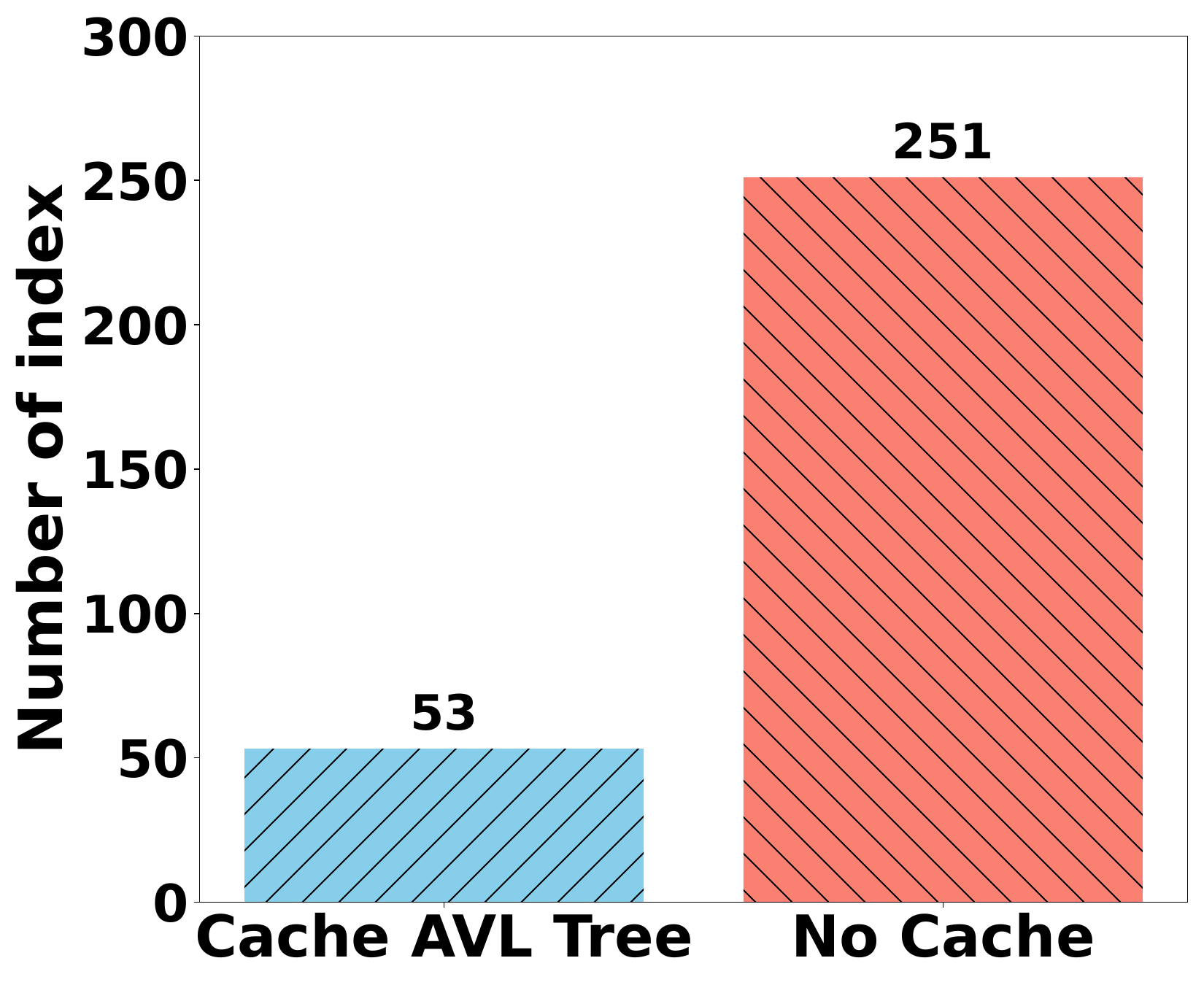}
        \caption{\small Storage Improvement.}
        \label{fig:Eff}
    \end{minipage}
\end{figure*}

\vspace{-5pt}
\subsection{Implementation}

\textbf{Blockchain Infrastructure.} We implement our framework using COSMOS \cite{kwon2019cosmos,belchior2021survey}, which enables interoperable blockchains with Tendermint consensus \cite{buchman2019latestgossipbftconsensus}, supporting governance, staking, and IBC.

\textbf{Datasets.} Subsets from WikiQA \cite{yang-etal-2015-wikiqa}, 
TruthfulQA \cite{lin2022truthfulqameasuringmodelsmimic}, and MathQA \cite{amini-etal-2019-mathqa} are used. 
We use a Query Generation Model (QGM) to produce query-answer pairs and embeddings, stored in a PostgreSQL database.

\textbf{Agent Simulation.} We simulate all blockchain agents  using Docker Compose. Transactions are communicated via Tendermint RPC. Full nodes include user, cache, and official validator nodes. Suppliers and validators operate as Tendermint validators. Users submit queries through \texttt{create-session} transactions; cache nodes retrieve from storage or forward to suppliers. Supplier nodes use LlamaIndex \cite{Liu_LlamaIndex_2022} to serve prompts. Validators score prompts, while the official validator applies cosine similarity to identify malicious behavior.

\textbf{Environment:} Ubuntu 24.04.1 LTS, Docker Compose v2.32.1, Python 3.10.16, Ollama v0.5.7, Ignite CLI v28.7.0, Cosmos SDK v0.50.11, CUDA 12.2, and RTX 4090 GPUs.

\vspace{-5pt}
\subsection{Experiment}
\textbf{Reputation Security:}
We evaluate the reputation mechanism proposed in Section~\ref{sc:reputation} by testing the performance of suppliers and validators separately. The setup includes eight honest nodes and three malicious nodes executing adversarial actions as defined in the threat model from Section~\ref{sc:secure}. The reputation of honest nodes is represented by the mean value and confidence interval(CI) curves. The results are shown in Fig.~\ref{fig:RP} for Knowledge Providers and Fig.~\ref{fig:RV} for Validators. The findings indicate that the reputation of malicious nodes decays to zero across all adversarial strategies, while honest nodes maintain stable and significantly higher reputations, validating the effectiveness of our reputation mechanism.

\textbf{Cache Performance:} We assess the performance of our proposed PROCache by comparing it with baseline LFU and LRU-k caching strategies. Fig.~\ref{fig:Cache} illustrates the performance metrics and the reputation of in-cache contents, which are indicative of service delay and quality closely related to Quality of Service (QoS). PROCache achieves comparable performance in delay reduction while enhancing the in-cache reputation. This demonstrates that our scheme effectively obsoletes low-quality content in the cache before it can cause widespread negative impact.

\textbf{Blockchain Storage:} Beyond enhancing transaction execution efficiency, PROCache significantly alleviates storage pressure on the blockchain ledger by approximately 80\%. We evaluate modifications to the cache's data structure by parsing the logs, as illustrated in Fig.~\ref{fig:Eff}. Our analysis reveals that out of 253 different questions and their variations, only 53 unique prompt indices, along with their content, are stored on the blockchain. This reduction is attributed to PROCache's ability to retrieve question variations directly and reuse historical answers from suppliers, eliminating the need to store duplicate question variations and their corresponding answers.

\section{Conclusion}
\label{Conclusion}

In this work, we propose BLOCKS, a novel blockchain-based knowledge-sharing framework that leverages the capabilities of \textit{Cosmos} and \textit{Tendermint} to enhance Large Language Model (LLM) services through secure and efficient integration of external, multi-domain knowledge. By addressing key challenges such as security, incentivization, and quality of service (QoS), BLOCKS facilitates robust collaboration among distributed knowledge providers. The incorporation of a reputation-based validation mechanism ensures the delivery of high-quality knowledge, fostering a reliable and trustworthy sharing environment. To further improve QoS, we introduce a novel knowledge caching mechanism that enables timely responses while maintaining both storage efficiency and content quality.

\bibliographystyle{IEEEtran}
\bibliography{IEEEabrv,ref}

\end{document}